# Sharp fall of electrical resistance for a small application of magnetic field on a metastable form of a compound, $Tb_5Si_3$, under pressure

Kartik K Iyer and E.V. Sampathkumaran[*]

*Tata Institute of Fundamental Research, Homi Bhabha Road, Colaba, Mumbai 400005, India*

**Abstract**

We report an unusual sensitivity of electrical resistivity ($\rho$) to an application of a small magnetic field in an intermetallic compound, $Tb_5Si_3$, under pressure. In this compound, there is a magnetic-field-induced first-order magnetic transition at 1.8 K. Under pressure, there is a metastable magnetic phase after reducing the field to zero. This metastable phase is relatively of higher $\rho$ and interestingly a small magnetic field (< 2 kOe) in the reverse direction results in a sharp fall of $\rho$ to restore virgin state $\rho$. The present finding could be relevant in spintronic applications.

[*]Electronic mail: sampath@mailhost.tifr.res.in

The magnetic materials undergoing magnetic transitions induced by magnetic field ($H$) have generated a lot of interest in recent years, for instance, from the angle of phase coexistence and metastability as well as of spintronic applications, where spin instead of (or in addition to) charge can be used. In such materials, e.g., in the most celebrated manganites [1], when the zero-field is attained after cycling through the metamagnetic transition field, there are certain situations in which the electrical resistivity ($\rho$) is dominated by the high-field magnetic phase (rather than by the initial state). When the magnetic field is reversed to negative direction, this 'high-field memory' still exists and it is not possible to restore the initial state unless the material is warmed above magnetic transition temperature and cooled again [see, for instance, Refs 2 - 3]. Here, we bring out an unusual situation in a metamagnetic compound, $Tb_5Si_3$ (under pressure at 1.8 K), in which the high-field memory of $\rho$ could be erased thereby restoring initial state resistivity, interestingly, by a small application of magnetic field (~ 2 kOe) in the reverse direction. This finding opens up an avenue for employing metastability for spintronic studies.

The compound under investigation has been known to form in $Mn_5Si_3$-type hexagonal structure (space group: $P6_3/mcm$) [4] and to undergo a complex helimagnetic ordering near ($T_N=$) 70 K [5]. Recently, we reported [6] that there is a field-induced first order magnetic transition near 60 kOe for $T$ = 1.8 K and the critical field ($H_{cr}$) at which this transition occurs decreases with increasing temperature. This transition becomes continuous as the temperature is raised to 5 K, but the hysteretic nature of the plots of magnetization ($M$) and magnetoresistance [defined as MR= $\{\rho(H)-\rho(0)\}/\rho(0)$] *versus H*, is retained till about 20 K, as though there is a tricritical point near this temperature. The most fascinating observation was that there is a sudden enhancement (by a few hundred percent) of MR *with the sign remaining positive*, in contrast to negative sign of MR expected across metamagnetic transitions. This anomalous MR observed in the entire temperature range in the magnetically ordered state is independent of whether the transition is discontinuous or continuous. While the exact origin of the huge positive MR at $H_{cr}$ was not clear, we speculate that the field-induced transition could arise from inverse metamagnetism, as demonstrated [7] for Co 3d itinerant magnetic system, $Er_{1-x}Y_xCo_2$, which means that paramagnetic fluctuations are induced at $H_{cr}$, at least at one of the two sites of Tb.

The polycrystalline sample was synthesized by arc melting together stoichiometric amounts of high-purity Tb (>99.9 wt%) and Si (>99.99 wt%) in an atmosphere of argon. The sample thus prepared was characterized by x-ray diffraction (Cu $K_\alpha$) and back scattered images of scanning electron microscope. The lattice constants (± 0.004 Å) obtained from the Rietveld fitting of the x-ray diffraction pattern assuming the occupation of Tb and Si at respective sites are: $a$= 8.441 and $c$= 6.347 Å. The $\rho$ and $M$ measurements were carried out employing a commercial cell (easy-Lab Technologies Ltd, UK). Commercial instruments (a physical property measurements system and a superconducting quantum interference device respectively, Quantum Design, USA) were employed for these measurements. The pressure studies were done in a hydrostatic pressure medium of Daphne oil in the case of magnetization studies (≤ 10 kbar) and a mixture of pentane and iso-pentane in the case of $\rho$ studies (≤15 kbar). All the measurements were done for the zero-field-cooled (from 250 K) conditions of the specimens. The electrical contacts of the leads to the specimen were made by a conducting silver paste.

In figure 1, we show the variation of MR as a function of $H$ at 1.8 K for $P$= 10 and 15 kbar. For the sake of comparison, the results at ambient pressure obtained in the earlier study [6] are also included. The curves were obtained after cooling the sample to 1.8 K in zero magnetic field. [It may be stated that $T_N$ was found to undergo a marginal decrease with increasing pressure]. The arrows and numericals placed near these arrows (shown for some cases only in figure 1) serve guides to the eyes. It was known [6] that, for ambient pressure, there is a sharp increase in MR at about 58 kOe in the virgin curve (see figure 1*a*) with an irreversibility. Though irreversibility of MR is observed in the pressure experiments as well (for instance, follow arrows 2 and 3 in figure 1*b*), the drop seen near 28 kOe while reversing the field towards zero for ambient pressure condition is absent for *P*= 10 and 15 kbar. Thus, under pressure, the irreversibility persists till *H*= 0, thereby expanding the hysteresis loop and making it 'open' in the first quadrant. *The point of central emphasis is that a small application of magnetic field (typically about 2 kOe) in the negative direction of H results in a dramatic fall of ρ restoring virgin state resistivity value (*follow arrow 4 in figure 1b*). It is thus interesting that the loop gets closed as soon as adjacent quadrant is entered.* On further ramping of magnetic field in the negative direction, we see the curve which is a mirror image of that in the positive-*H* quadrant. It may be noted that a small positive value of *H* after traversing through the cycle in the negative-*H* quadrant needs to be applied to restore initial state (follow the paths shown by arrows 4, 5, 6, 7, 8 in figure 1*b*). Such features were observed even in subsequent field-cyclings. Similar curves are obtained for 15 kbar. The low-field behavior is shown in an expanded form as an inset, for instance for 10 kbar in the figure 1*b* to highlight this observation.

Such a behavior in MR-*H* plot is absent for a marginal increase of temperature, say for 5 K as shown in figure 1*e* and 1*f*. At this temperature and at a pressure of 10 kbar, the value of $H_{cr}$ in the forward cycle is found to be reduced marginally compared to that at ambient pressure (compare the curves in figures 1*d* and 1*e*). However, the reverse transition is complete well-within the first quadrant itself for this temperature. But the field at which this occurs is reduced by about 12 kOe compared to that at ambient pressure. Thus, there is a clear evidence for the expansion of hysteresis loop under a pressure of 10 kbar (consistent with the trend at 1.8 K under pressure). At a further higher pressure (15 kbar) at 5 K, the 'reverse' $H_{cr}$ is pushed down further; it appears that the transformation to initial state is partial as indicated by an intermediate value of the MR as $H \to 0$. This 'butterfly'-shaped behavior of MR(*H*) curve of this partial conversion [See, for instance, references 2, 3, 8, 9], which is addressed further below, can be clearly seen. The virgin curve lies distinctly outside the hysteresis loop, which is a signature of first-order transitions.

At this stage, it is quite instructive and important to recall that, among manganites [2, 8, 10] and intermetallics [11, 12], the hysteresis loop actually broadens with decreasing temperature. The variation of (lower) $H_{cr}$ for the reverse cycle with temperature is non-monotonic (showing a peak in its plot as a function of *T*) whereas $H_{cr}$ in the forward cycle decreases monotonically with increasing *T*. In recent years, this non-monotonic behavior of lower-$H_{cr}$ is taken as a signature [8, 12] for the arrest of complete transformation of the 'supercooled' phase to initial phase. This results in a mixture of initial and high-field phases for some field-range. This phase co-existence can even persist as $H \to 0$ which manifests itself as an open loop in the first quadrant of *M(H)* or MR(*H*). In the present compound, a careful look at the figures 2 and 3 in Ref. 6 reveals that the hysteresis loop behaviors are similar to that reported for other materials in the literature in the sense that $H_{cr}$ in the down cycle undergoes a non-monotonic variation with *T*. Therefore, in the discussions, we keep the phase co-existence phenomenon in mind.

We now look at the isothermal magnetization under a pressure of 10 kbar at 1.8 and 5 K (figure 2) for a comparison with MR behavior. In figure 2, we have included the data under ambient pressure presented in Ref. 6. At ambient pressure, at both 1.8 and 5 K, irreversibility is apparent, but the transformation to the initial state is complete well before $H$ is reduced to zero. This is found to be the case for -$H$ direction as well, as shown in figure 2. These are in agreement with MR behavior. With respect to the influence of pressure, we first look at the behavior at 5 K: The sudden (but continuous) increase of $M$ near 58 kOe persists while ramping the field upwards, but the transition while reducing the field is apparently absent and $M$ decreases essentially linearly with $H$. $M$ gets gradually smaller, attaining almost zero value at $H=0$. It must be stressed that no sharp jump in $M$ could be observed for small applications of $H$ in the opposite direction, in contrast to that seen in MR($H$) curves. These observations could be consistently interpreted in terms of the paramagnetic nature of the supercooled high-field phase whose remnant magnetization should tend to zero as H$\rightarrow$0, like that of initial (antiferromagnetic-like) phase. Hence it is quite natural that remnant magnetization does not offer a way to find the relative fractions of both the phases. We elaborate on this point further below.

If one views the field-induced transition within the conventional 'metamagnetism', the absence of any sharp transition in the 'down' field cycle in the high pressure $M(H)$ curve tempts one to propose that the transformation of the high-field ferromagnetic phase is gradual with decreasing field and only a small fraction of this phase continues to persist down to $H=0$ (which can be inferred from the negligibly small remnant $M$ value). In general, if the high-field phase is relatively highly conductive than the initial phase and if the fraction of the high-field phase is sufficient enough to allow a percolative path, then the conductivity is typical of high-field phase only as demonstrated in many systems [see, for examples, references 2, 3, 9, 11]. However, if one estimates the fractions from the remnant $M$ value in the present high pressure studies, it is clear that the initial phase (almost 100%) should dominate as $H\rightarrow 0$; then the MR curve should have fallen back on the virgin curve well-within the first quadrant. This could possibly explain the fall of MR below about 25 kbar at 5 K for $P=10$ kbar. At this temperature, as the pressure is increased to 15 kbar, the peak field in $MR(H)$ plot is shifted (from 25 kOe at 10 kbar) to nearly 12 kOe, as though the high-field fraction tends to increase at a higher pressure. An 'intermediate' MR value (in other words, 'butterfly' behavior of MR curve) is observed as $H$ is reversed to zero for this pressure, which appears to be consistent with the above picture. Accordingly, one should have seen 'enhanced' remnant $M$ value, as demonstated, for instance, for $Nd_7Rh_3$ (see figure 2 in Ref. 11). Our pressure cell does not permit us to perform studies at 15 kbar to verify whether the remanent $M$ is truly enhanced. However, if one extends the above line of arguments to $M$-$H$ behavior at 1.8 K under pressure (figure 1$b$), then there is some inconsistency in the above explanation considering 'metamagnetism': That is, one has to assume that the fraction of high-field phase in zero-field after a field-cycling is extremely large, particularly noting that the 'supercooled' high-field phase is more resistive than the initial phase and the value of $\rho$ after returning the field to zero is very large. In other words, a marginal lowering of temperature (from 5 K to 1.8 K) is equivalent to an increase of pressure above 15 kbar at 5 K. Within the idea of conventional 'metamagnetism', such a large fraction of high-field phase should have dramatically enhanced the magnitude of remnant $M$ comparable to the value of $M$ beyond $H_{cr}$. In sharp contrast to this expectation, negligibly small value at 10 kbar at this temperature is observed. However, if one invokes the idea of 'inverse metamagnetism', then the observed small remnant $M$ and large $\rho$ can be consistently explainable, as $M$ for both 'supercooled' paramagnetic and initial antiferromagnetic phase should be close to zero for zero field. In support of possible paramagnetic phase, the $\rho$ as $H\rightarrow 0$ varies quadratically with $H$ in figure 1*b* and 1*c*.  Incidentally, there is a gradual fall of MR beyond $H_{cr}$ and MR even passes through zero near 90 kOe in high pressure experiments (see figure 1); under ambient pressure conditions, the same trend of MR is observed at high fields, however with the sign remaining positive near 100 kOe (see Ref. 6).   This gradual fall may also be attributed to the suppression of paramagnetic fluctuations by magnetic field.  Irrespective of the nature of the phases as $H \rightarrow 0$ (as well as beyond $H_{cr}$), it is fascinating that a small field in the opposite direction is enough to restore the electrical conduction typical of the initial phase. This is a mystery at present.

Summarizing, in the pressurized form an intermetallic compound, $Tb_5Si_3$, high electrical resistivity of a metastable phase obtained by field-cycling across first-order field-induced magnetic transition can be switched to a low-resistive state characteristic of the virgin phase, by a small application of magnetic field in the opposite direction.  It is worthwhile to explore this principle for spin-valve based devices employing bulk materials. The present finding is another demonstration of effects due to metastability in systems exhibiting first-order transition.

**References**

1. See, for a review, E. Dagotto, T. Hotta, and A. Moreo, *Phys. Rep.* **344** (2001).
2. Y. Tokura, H. Kuwahara, Y. Moritomo, Y. Tomioka, and A. Asamitsu, *Phys. Rev. Lett.* **76**, 3184 (1996).
3. T. Kimura, Y. Tomioka, R. Kumai, Y. Okimoto, Y. Tokura, *Phys. Rev. Lett.* **83,** 3940 (1999).
4. K.S.V.L. Narasimhan, H. Steinfink, and E.V. Ganapathy, *J. Appl. Phys.* **40,** 51 (1969).
5. I.P. Semitelou, H. Konguetsof, and J.K. Yakinthos, *J. Magn. Magn. Mater.* **79,** 131 (1989).
6. S. Narayana Jammalamadaka, Niharika Mohapatra, Sitikantha D. Das, and E.V. Sampathkumaran, *Phys. Rev. B* **79,** 060403R (2009).
7. R. Hauser, E. Bauer, E. Gratz, H. Müller, M. Rotter, H. Michor, G. Hilscher, A.S. Markosyan, K. Kamishima, and T. Goto, *Phys. Rev. B* **61**, 1198 (2000).
8. R. Rawat, K. Mukherjee, K. Kumar, A. Banerjee and P. Chaddah, J. Phys.: Condens. Matter. **19,** 256211 (2007).
9. M.A. Manekar, S. Chaudhary, M.K. Chattopadhyay, K.J. Singh, S.B. Roy, and P. Chaddah, *Phys. Rev. B* **64**, 104416 (2001).
10. H. Kuwahara, Y. Tomioka, A. Asamitsu, Y. Morimoto, and Y. Tokura, *Science* **270**, 961-963 (1995).
11. K. Sengupta and E.V. Sampathkumaran, *Phys. Rev. B* **73**, 020406(R) (2006).
12. P. Kushwaha, R. Rawat, and P. Chaddah, *J. Phys.: Condens. Matter* **20**, 022204 1-7 (2008).

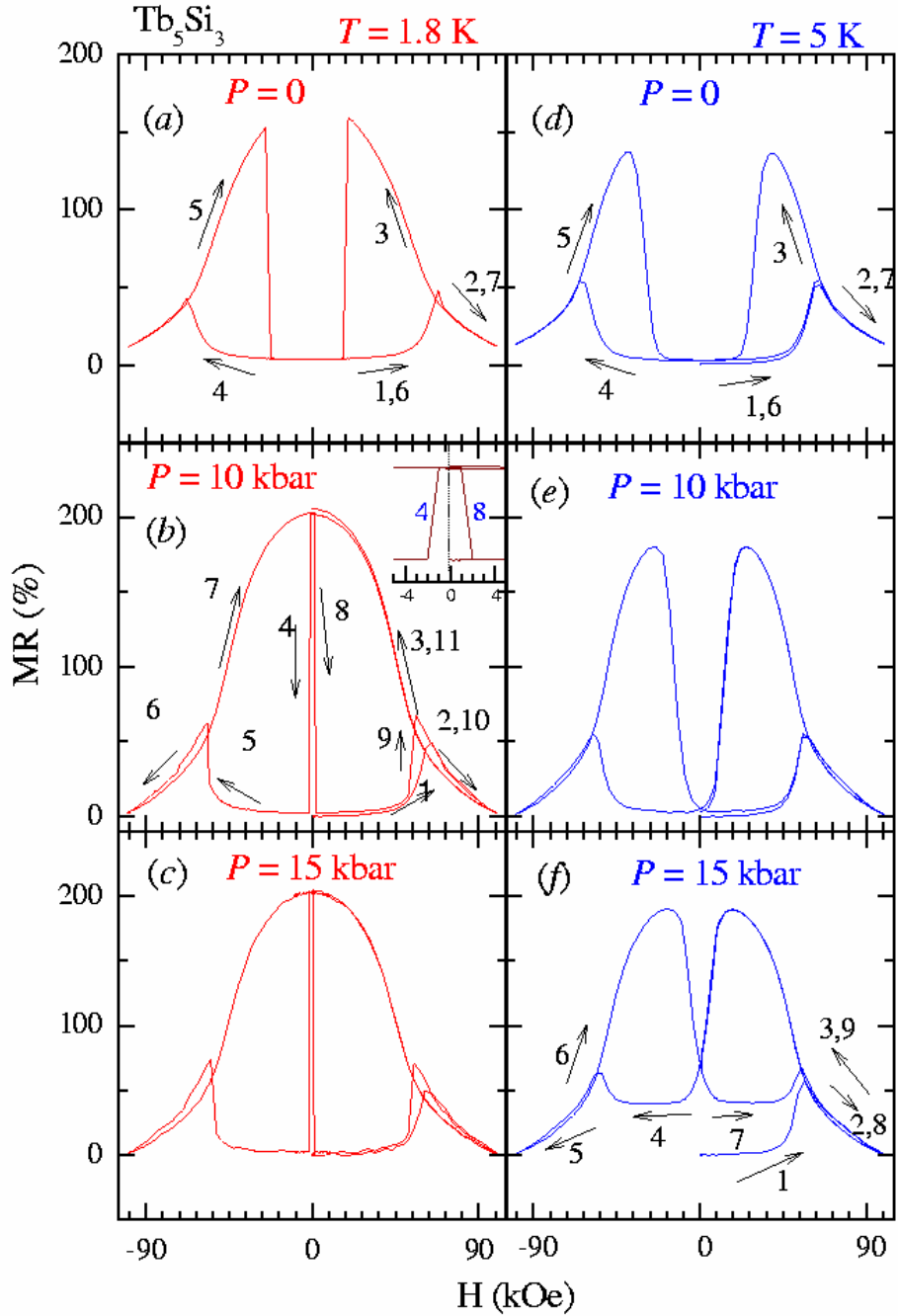

Figure 1:
Magnetoresistance, defined as $[\rho(H)-\rho(0)]/\rho(0)$, as a function of magnetic field for $Tb_5Si_3$ under pressure (10 and 15 kbar) at 1.8 and 5 K. The arrows and numbers placed near the curves (shown typically for some graphs only) serve as guides to the eyes. The low-field region is shown in an expanded form for $T$= 1.8 K to highlight the sharp fall while switching the magnetic field direction.

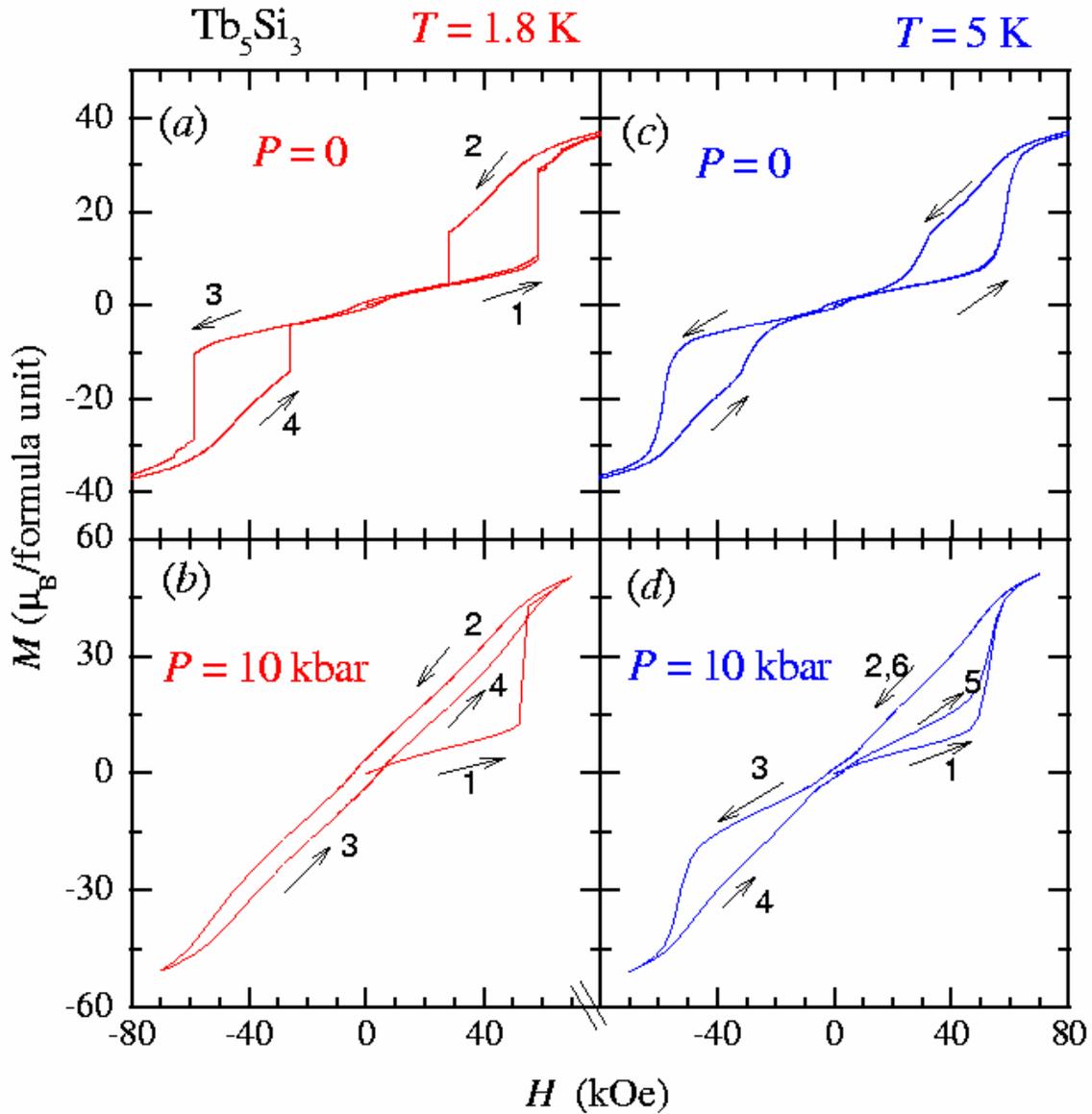

Figure 2:
(color online) Isothermal magnetization behavior at 1.8 and 5 K for $Tb_5Si_3$ under ambient pressure and 10 kbar. The arrows and the numbers placed serve as guides to the eyes.